\documentclass[a4paper,final,titlepage,10pt]{iopart}
\usepackage{epsfig,epstopdf}
\usepackage{hyperref}
\usepackage{cuted}
\hypersetup{
    breaklinks=true
    bookmarks=true,         
    unicode=false,          
    pdftoolbar=true,        
    pdfmenubar=true,        
    pdffitwindow=false,     
    pdfstartview={FitH},    
    pdftitle={Many-qubit quantum state transfer via spin chains},    
    pdfauthor={T. J. G. Apollaro et al},     
    pdfsubject={Quantum information processing},   
    pdfcreator={Anto},   
    pdfproducer={CiccioP}, 
    pdfkeywords={Quantum state transfer} {Teleportation} {Spin-Chains}, 
    pdfnewwindow=true,      
    colorlinks=true,        
    linkcolor=blue,          
    urlcolor=blue,        
    filecolor=red,      
    citecolor=blue     
}
\usepackage{amssymb,bbm,amsfonts,iopams,subfigure}
\newcommand{\bra}[1]{\left< #1 \right|}

\newcommand{\ket}[1]{\left| #1 \right>}

\newcommand{\text}[1]{\mathrm{#1}}
\newcommand{\fourIdx}[5]{
\setbox1=\hbox{\ensuremath{^{#1}}}
\setbox2=\hbox{\ensuremath{_{#2}}}
\setbox5=\hbox{\ensuremath{#5}}
\hspace{\ifnum\wd1>\wd2\wd1\else\wd2\fi}
\ensuremath{\copy5^{\hspace{-\wd1}\hspace{-\wd5}#1\hspace{\wd5}#3}_{\hspace{-\wd2}\hspace{-\wd5}#2\hspace{\wd5}#4}}
}

\newcommand{\antoC}[1]{#1}

\begin{document}
\title{Many-qubit quantum state transfer via spin chains}
\author{T.~J.~G. Apollaro$^{1,2,3}$}
\ead{tony.apollaro@fis.unical.it}
\author{S.~Lorenzo$^{4}$}
\ead{salvatore.lorenzo@unipa.it}
\author{A.~Sindona$^{1,2}$}
\ead{sindona@fis.unical.it}
\author{S.~Paganelli$^{5}$}
\ead{pascualox@gmail.com}
\author{G.~L.~Giorgi$^{6}$}
\ead{gianluca0giorgi@gmail.com}
\author{F.~Plastina$^{1,2}$}
\ead{francesco.plastina@fis.unical.it}
\vskip 12pt
\address{$^1$Dipartimento di Fisica, Universit\`a della Calabria, 87036 Arcavacata di Rende (CS), Italy}
\address{$^2$INFN sezione LNF-Gruppo collegato di Cosenza, Italy}
\address{$^3$Centre for Theoretical Atomic, Molecular, and Optical Physics, School of Mathematics and Physics, Queen's University Belfast,BT7,1NN, United Kingdom}
\address{$^4$Dipartimento di Fisica e Chimica, Universit\`{a} degli Studi di Palermo, via Archirafi 36, I-90123 Palermo, Italy}
\address{$^5$International Institute of Physics, Universidade Federal do Rio Grande do Norte, 59012-970 Natal, Brazil}
\address{$^6$INRIM, Strada delle Cacce 91, I-10135 Torino, Italy}

\begin{abstract}
The transfer of an unknown quantum state, from a sender to a
receiver, is one of \antoC{the main requirements to perform quantum
information processing tasks.
In this respect, the state transfer} of a single qubit by means of spin chains
has been widely discussed, and many protocols aiming at performing this task
have been proposed.
Nevertheless, the \antoC{state transfer} of more than one
qubit has not been properly \antoC{addressed so far}.
In this paper, \antoC{we present} a modified version of a recently proposed \antoC{quantum state transfer} protocol~[Phys. Rev.  A {\bf{87}}, 062309 (2013)] to obtain a quantum channel for the \antoC{transfer of two qubits}.
This \antoC{goal} is achieved by exploiting Rabi-like
oscillations \antoC{due to} excitations induced by means of strong and
localized magnetic fields.
We derive exact analytical \antoC{formulae} for
the fidelity of the \antoC{quantum state transfer,} and obtain a high-quality transfer for
general quantum states as well as for specific classes of states
relevant for quantum information processing.
\end{abstract}

\date{\today}

\maketitle
\twocolumn
\section{Introduction}\label{S.Introduction}

Quantum Information Processing~(QIP) is a fundamental resource for
the next generation of technological devices based on quantum
principles.
The field of QIP is attracting a continuously
increasing interest since its birth, a few decades ago, and branches
\antoC{into several subfields}: quantum computation, quantum
communication, quantum algorithms, and quantum cryptography, just
to name a few~\cite{BooksQIP}.
\antoC{Possible broad-reaching
applications of such devices are triggering experimental and theoretical
efforts, which aim at controlling and manipulating the systems on a
quantum scale and characterizing their many-body
properties.}

\begin{figure}[h]
\includegraphics[width=\columnwidth]{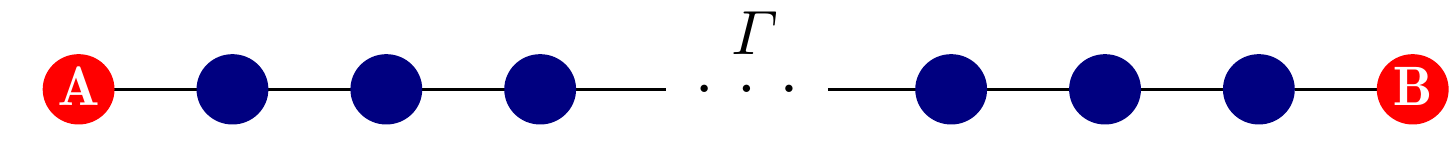}
\caption{(Color online) The quantum state $\ket{\Psi(0)}_1$,
encoded on spin $A$, \antoC{and} residing on the first site of a spin chain,
\antoC{is meant to be transferred} to spin $B$, \antoC{which resides} on site $N$, at some time $t^*$, by exploiting the coherent dynamics of the
quantum channel $\Gamma$.
\label{F.1qubitQSTQED}}
\end{figure}

In this paper, we focus on Quantum State Transfer~(QST), perhaps
the simplest communication protocol, \antoC{designed to} send a
quantum state \antoC{through} interconnected
on-chip architectures\antoC{, from one node to another}.
\antoC{QST} can be useful \antoC{in} short distance
communication\antoC{s} to avoid interfacing problems with flying
qubits. In its simplest version, a QST task is performed between parties
connected \antoC{via} quantum spin channels, that is, $1$D
interacting spin-$\frac{1}{2}$ chains working as data buses, see
Fig.~\ref{F.1qubitQSTQED}.

Since the seminal paper \antoC{by} Bose~\cite{Bose03}, \antoC{numerous} protocols
acting on spin-$\frac{1}{2}$ chains have been proposed and investigated in
order to achieve a high QST fidelity. \antoC{These can be classified into two broad groups, namely} time-dependent and time-independent
\antoC{protocols}~(see Refs.~\cite{channels} and references
therein for review). In the former protocols, the transfer operation is
performed by a time-modulation of the interaction parameters of
the spin model; whereas, in the latter scenario, the interaction
parameters are kept fixed during the execution of the QST
protocol.
Notwithstanding the rapid improvement of the ability to
control the interactions between quantum spins~(especially in cold
atom set-ups~\cite{Antonello}), the accuracy required to perform time-dependent
protocols seems to be \antoC{still} out of reach.
Hence, in this paper, we
will focus \antoC{on} time-independent QST protocols.

As far as the transfer of the quantum state of a single qubit is
concerned, it has been shown~\cite{Christandletal05} that
perfect-QST and perfect entanglement distribution are completely
equivalent. \antoC{In other words,} if a quantum channel is capable of
transferring a quantum state with unit fidelity from a sender to a
receiver, then it \antoC{will also allow} for a maximally entangled state to
be distributed between these two parties.
A high-fidelity QST data
bus can be employed as an entangling gate between two
spins~\cite{Banchietal11PRL}, producing, e.g., any one of the
maximally entangled Bell states \antoC{with two qubits:}
$\ket{\Psi^{\pm}}{=}\frac{1}{\sqrt{2}}\left(\ket{01}{\pm}\ket{10}\right)$
and
$\ket{\Phi^{\pm}}{=}\frac{1}{\sqrt{2}}\left(\ket{00}{\pm}\ket{11}\right)$,
here expressed in the (logical)~basis of eigenstates of the Pauli
matrix $\sigma^z$.
The shared entangled state can be then
exploited, via the celebrated teleportation protocol~(TP)~\cite{Bennetetal93},
as a resource to teleport an
unknown quantum state from one location to another, \antoC{thus achieving}
the desired QST.
The TP requires the sender to perform a two-qubit
\antoC{measurement}, followed by the transmission of two bits of classical
communication to the receiver\antoC{, which report the result of the measurement.
F}inally, a conditional unitary operation on the
target spin is carried out by the receiver himself.
As a consequence,
one-qubit QST can be achieved deterministically  by the use of a
two-qubit Bell state teleportation channel.

Unfortunately, there is not such a clear-cut equivalence when
multipartite QST is addressed, that is, when the quantum state to
be transferred \antoC{has} $n\geq2$ qubits. The reason for this
difficulty in connecting the $n$-QST with
multipartite entangled states, \antoC{which would be} useful as a resource for
implementing a teleportation channel, is mainly due to the
in-equivalence of entangled states in the multipartite regime.
For
instance, the entanglement of teleportation
$E_T$~\cite{Rigolin05} \antoC{has been adopted} as a quantifier of the capability of an
entangled state to act as a teleportation channel. It turned out
that 4-qubits GHZ-states,
$\ket{GHZ}=\frac{1}{2}\left(\ket{0000}+\ket{1111}\right)$, have
$E_T=\frac{1}{2}$ and only special classes of 2-qubit states can be
deterministically teleported~\cite{Lee01}.
On the other hand, a different class of maximally entangled 4-qubit entangled
states, the so-called W-states,
$\ket{W}=\frac{1}{2}\left(\ket{0001}+\ket{0100}+\ket{0100}+\ket{1000}\right)$,
\antoC{have} $E_T=0$, and no deterministic 2-QST is possible.
\antoC{Actually,}
it is possible to
teleport a 2-qubit state by using a 4-qubit entangled state~\cite{LeeMO02}, and
an explicit protocol for generic $n$-QST
has been introduced in terms of the so-called $2n$ generalized
Bell states~\cite{Rigolin05}, which, indeed, have unit teleportation entanglement.

\begin{figure}[h]
\includegraphics[width=\columnwidth]{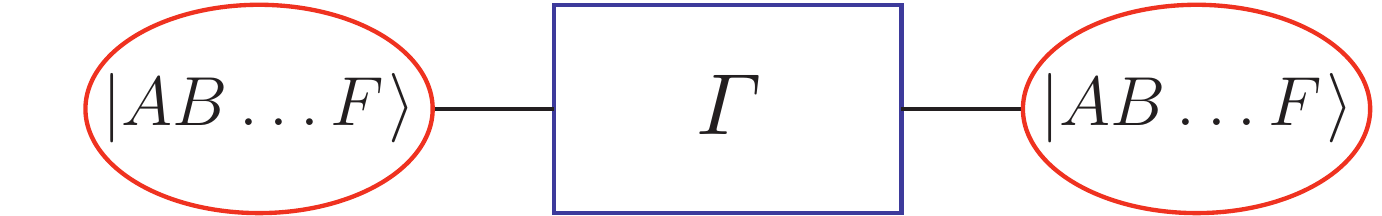}
\caption{(Color online) The quantum state $\ket{AB...F}$ of the
sender qubits  is aimed \antoC{at being} transferred to the receiver spins
via the quantum channel $\Gamma$.
\label{F.NqubitQSTA}}
\end{figure}

Notwithstanding the possibility to teleport an $n$-qubit state by
means of suitable entangled \antoC{states,} shared by the sender and the
receiver, in this paper we address the question in terms of the
{\textit{transport}} of the state.
This presents some advantages
with respect to the use of a teleportation channel: first of all,
in order to perform the $n$-QST via TP, there is the need to
implement generalized Bell \antoC{measurements, which} up to now
seem to be highly non-trivial; furthermore, sender and receiver
\antoC{need} to protect efficiently their shared pure entangled state from
the environment, as the entanglement of teleportation has to be
\antoC{unity} 
in order to achieve a successful TP.
If this is not the case,
little can be said\antoC{. Indeed,} to be best of our knowledge, an analysis
of $n$-QST via TPs employing mixed states has yet to be performed,
and not much is known about \antoC{the efficiency of such tasks}.
Therefore, we
consider the case of QST in a setting where the state is encoded
at one end of the spin chain\antoC{. Then,} by exploiting the natural
dynamics of the quantum channel, we aim at retrieving the same
state~(up to some unitaries) at the other end, as shown
schematically in Fig.~\ref{F.NqubitQSTA}.

The paper is organized as follows: in Sec.~\ref{S.Overview} we
review some of the main single-qubit QST mechanisms \antoC{based on}
spin chains; in Sec.~\ref{S.Dynamics} we solve the many-body
dynamics for our model, which will be \antoC{used} in
Sec.~\ref{S.Results} to present our main results on the QST
fidelity. Finally, in Sec.~\ref{S.Conclusions} we draw \antoC{some}
conclusions.

\section{Overview of single-qubit quantum state transfer protocols}\label{S.Overview}

In this Section, we will focus on $1$-QST \antoC{as} performed by means of an
open \antoC{and} finite spin-$\frac{1}{2}$ chain\antoC{, like the one depicted} in
Fig.~\ref{F.1qubitQSTQED}.  The Hamiltonian describing its
dynamics is taken \antoC{to be} of the $XX$-Heisenberg type, with
nearest-neighbour interactions only\antoC{, plus} a magnetic field along the
$z$-axis:
\begin{eqnarray}\label{E.HamBus}
H&=&-\sum_{l{=}1}^N J_l(\sigma^x_l \sigma^x_{l+1}+\sigma^y_l \sigma^y_{l+1})+\sum_{l{=}1}^N h_l\sigma^z_l~.
\end{eqnarray}
The initial state, encoded on the sender spin $A$, reads
$\ket{\psi(0)}=a\ket{0}+b\ket{1}$, with $|a|^2+|b|^2=1$, whereas
the rest of the chain, including the quantum channel $\Gamma$ and the receiver $B$, is initialized in
$\ket{\Gamma}\!\ket{B}=\otimes_{j{=}2}^{N}\ket{0}_j$.
The
evolution of the overall state is
$\ket{\Psi(t)}=a\otimes_{j{=}1}^{N}\ket{0}_j+b\, e^{- i H t}
\ket{1}_1\otimes_{j{=}2}^{N}\ket{0}_j$.
By tracing out all of the
spins but $B$, one obtains the~(generally mixed) state of the
receiver spin \antoC{on site $N$:} $\rho^B(t)=Tr_{\neq
N}\left(\ket{\Psi(t)}\!\!\bra{\Psi(t)} \right)$.
The fidelity
between the state transferred to the receiver \antoC{on site} $N$, and the
state encoded initially on spin $1$ by the sender, is given
by~\cite{Josza94}:
\begin{equation}\label{E.Fidelity}
F\left(\ket{\psi(0)}\!\!\bra{\psi(0)}_A,\rho_B(t)\right)
{=}\sqrt{{}_A\!\bra{\psi(0)}\rho_B(t)\ket{\psi(0)}_A}~.
\end{equation}

The quality of a QST-bus, however, cannot \antoC{be simply} evaluated by
considering the fidelity of the  transfer of a single, specific
input state, but rather by an average QST-fidelity obtained over
some \antoC{classes} of states.
We will consider \antoC{here} the average fidelity
$\bar{F}(t)$, evaluated by \antoC{integration} over the Bloch sphere of all
possible pure input states, as a QST figure of merit.
In
Ref.~\cite{Bose03} it has been shown that the average fidelity is
given~(up to a phase that can be adjusted by \antoC{a suitable} magnetic
field) by the expression
\begin{equation}\label{E.Fid01}
\bar{F}(t)=\frac{1}{2}+\frac{|f_{N1}(t)|}{3}+\frac{|f_{N1}(t)|^2}{6}~,
\end{equation}
where $f_{N1}(t)$ is the transition amplitude of the excitation~(or, equivalently, of the spin-flipped state $\ket{1}$) from site $1$ to site $N$.
The total magnetization along the $z$-axis
commutes with the hamiltonian given by Eq.~\ref{E.HamBus},
\antoC{then, the} subspaces with a fixed number of excitations are invariant under
its action. As a consequence, the amplitude is
straightforwardly evaluated via
\begin{equation}\label{E.fRS}
f_{N1}(t)= \bra{N}e^{- i H t} \ket{1}=
\sum_{k{=}1}^{N}\bra{N}\mathbf{a}_k\rangle \! \langle\mathbf{a}_k\ket{1} e^{-i \lambda_k t}~,
\end{equation}
where $\{\lambda_k,\ket{\mathbf{a}_k}\}$ are the single-particle
eigenvalues and eigenvectors of Eq.~\ref{E.HamBus}.
\antoC{Now, the} average fidelity $\bar{F}(t)$ depends~(\antoC{monotonically})
only on the modulus of the  excitation transition amplitude \antoC{$|f_{N1}(t)|$}
from site $1$ to site $N$. \antoC{Besides,}
$f_{N1}(t)$~(Eq.~\ref{E.fRS}) is a sum of products of the overlaps of any (single
particle) eigenstate with the initial and final states $\ket{1}$
and $\ket{N}$. \antoC{Each these overlaps} bring in a phase
factor determined by the eigenvalues $\lambda_k$ and \antoC{the $k$-sum in Eq.~\ref{E.fRS}}
runs over all \antoC{the eigenstates of Eq.~\ref{E.HamBus}. It} turns out that the different
components \antoC{of $f_{N1}(t)$} interfere destructively. \antoC{Accordingly}, a very low quality of
1-QST is obtained for uniformly coupled spins in long chains,
unless \antoC{a specific} procedure~\cite{channels} is applied to single-out some of the terms
\antoC{in the sum given in Eq.~\ref{E.fRS}}.

From Eq.~\ref{E.Fid01} it is evident that, as far as $1$-QST is
concerned, high-fidelity protocols aim at maximizing the transfer
of the spin-flipped state $\ket{1}$ from site $1$ to site $N$.
\antoC{In order to maximize 
$f_{N1}(t)$, \antoC{within time-independent hamiltonian protocols,}
several strategies have
been adopted
, which can be
broadly classified in:} a.) couplings-engineering
methods~\cite{Christandletal04, DiFrancoPK08, Kay10,
MarkiewiczW09}, b.) ballistic transfer~\cite{Banchietal10,
Banchietal11, Apollaroetal12, ZwickO11, Banchi13,Paganellietal09}, and c.)
Rabi-like dynamics~\cite{PaganellidPG06, Lorenzoetal13, Yaoetal11,
Wojciketal05,  GiorgiB,pla07,dePasquale05}.

In protocols relying on methods \antoC{of the a.)-type,} the nearest-neighbor spin
couplings are chosen in such a way that the energy spectrum
becomes linear, \antoC{thus allowing a} dispersion-less transfer of the
excitation; an instance of such a coupling set \antoC{is given by}
$J_n=\sqrt{n\left(N-n\right)}$, which yields a perfect QST, that is
$F=1$. With regards to ballistic transfer settings of \antoC{the b.)-type}, the main
idea is to modify \antoC{only} the couplings of the sender and the
receiver to the rest of the spin chain\antoC{.
These couplings are chosen} in such a way that,
even tough the overall energy spectrum is non linear, most of the modes
\antoC{prominently} involved in the dynamics reside in \antoC{a portion} of the spectrum\antoC{, which is} approximately linear. For example, by tuning
$J_1=J_{N-1}= c\, N^{-\frac{1}{6}}$~(\antoC{with $c$ reading} $c=1.030$
for $N\gg 1$) average fidelities higher than $99\%$ are
achieved~\cite{Banchietal11}; furthermore, allowing also for
modulations of the second and last-but-one couplings,
$\bar{F}$ \antoC{gets} as high as $99.9\%$ \antoC{in} arbitrarily long
chains~\cite{Apollaroetal12}. Finally, methods of \antoC{c.)-type} consist
in restricting the dynamics to just two~(or three) modes of the
spectrum in such a way that effective Rabi oscillations of the
excitations take place between the endpoints of the chain.
\begin{figure}[h]
\includegraphics[width=\columnwidth]{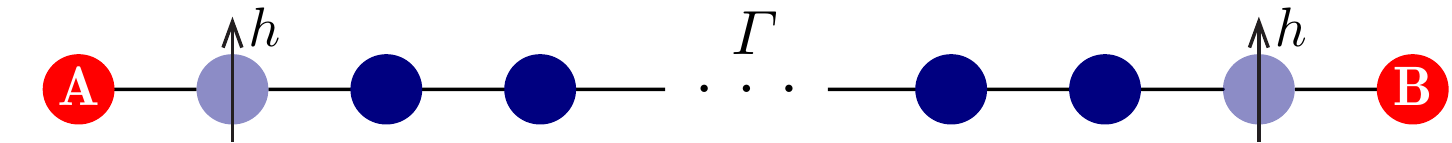}
\caption{(Color online) The state of qubit A is transferred to
qubit B by means of Rabi-like oscillations between eigenstates
localized on A and B, due to a strong magnetic field on the
neighboring ``barrier'' spins.}
  \label{F.1qubitQSTA}
\end{figure}

Up to now, these schemes have been applied mainly to the QST of a
single qubit~(even if this does not mean, in general, that the
dynamics is restricted to the single excitation subspace, as the
chain could be initialized in a state different  from
$\ket{0}^{\otimes N}$,~\cite{Wangetal12}). However, it would be of
uttermost importance to have a {\textit{single}} channel able to
perform different QIP tasks, \antoC{such as}, for instance, a QST of
arbitrary $n$-qubits or a QST \antoC{involving} more than just a single sender
and receiver.

In the next section we will focus on a specific Rabi-like 1-QST
protocol~(Fig.~\ref{F.1qubitQSTA}), \antoC{presented} in
Ref.\cite{Lorenzoetal13}, \antoC{with the idea} to investigate its capability
to be used as an $n$-QST bus. \antoC{The relevant feature of the protocol} is given by the
presence of (strong) magnetic fields applied only on the spins
sitting at sites 2 and $N-1$, whereas the spin-spin coupling is
assumed to be uniform. These strong magnetic fields on the
\antoC{next-neighboring spins} to the sender and the receiver, named hereafter
{\textit{barrier}} spins, play \antoC{a crucial} role \antoC{in} effectively decoupling
the end-spins from the rest of the chain. \antoC{Then,} an
effective Rabi-like hamiltonian $H_R\simeq\omega_1
\ket{\Psi_1}\!\!\bra{\Psi_1}+\omega_2
\ket{\Psi_2}\!\!\bra{\Psi_2}$ \antoC{can be written down}, with the two modes $\ket{\Psi_i}$
bi-localized on sender and receiver: $\ket{\Psi_{1,2}}\simeq
\frac{1}{\sqrt{2}}\left(\ket{01}\pm\ket{10}\right)_{1N}$. \antoC{In the following,} we will
 investigate if such a scheme succeeds also in the $n$-QST scenario.

It is worth mentioning that such a setup has been already extended
in Ref.~\cite{Paganellietal13} to perform a one-to-many routing
protocol.

\section{Dynamics in the many-body regime}\label{S.Dynamics}
In the usual single qubit QST scenario~(sketched in
Fig.~\ref{F.1qubitQSTQED}), a pure state $\ket{\Psi}$ is encoded
at time $t=0$ in a spin located at site $1$~(the sender)\antoC{. The
aim} is to retrieve, at some time $t=t^*$, the very same state in a
spin located at site $N$ (the receiver), via
the natural~(coherent) dynamics of the channel ruled by $H$.
As the number of
excitations (flipped spins) remains constant during the evolution,
the dynamics takes place in the single excitation subspace, as
witnessed by Eqs.~\ref{E.Fid01} and~\ref{E.fRS}. Though, one of the
applications of the QST protocol is, for instance, to transfer the
output of a QIP task, performed by a quantum processor, to another
quantum device, and often it is the case that such an output
consists of more than just one qubit. As a consequence, the output
has to be encoded in a larger Hilbert space, say, that of $n$
spins; so the resulting dynamics of the $n$-QST may
take place simultaneously in all the subspaces with $m\leq n$
excitations.
These components of the initial state have to evolve
in such a way that there exists a certain time $t^*$ at which the
initial state is rebuilt at the $n$ receiver spins.
It is evident
that not only the analytical, or numerical, complexity of the
problem increases exponentially with the number of spin flipped in
the initial state, but also the achievement of high-quality
transfer may become considerably more difficult w.r.t. the 1-QST
case.

One way to circumvent the problem would be to employ $n$
{\textit{identical}}, non-interacting quantum channels, each
transferring one component of the state~\cite{Christandletal05}.
\antoC{However,} there are some drawbacks in this configuration: all of the $n$
channels have to posses exactly the same technical specifics, as
each of the component of the whole $n$-qubit quantum state has to
be delivered at the same time $t=t^*$ to the corresponding
receiver.
\antoC{Unfortunately,} experimental imperfections \antoC{are likely to} modify the coupling
strength between spins in the different chains, thus yielding
different optimal transfer times $t^*$. These times could also
depend on the state to be transferred, thus making the $n$-qubit
QST a quite involved task in the presence of disorder; see, e.g.,
Refs.~\cite{Anania}. Furthermore, in order to fulfil the
assumption of coherent dynamics, the $n$ quantum channels have to
be efficiently protected from detrimental environmental effects,
which may result in a demanding technical (and economical)
request. Also endowing the QIP devices by moving parts give rise
to easily presumable difficulties. It would be therefore
interesting to explore the possibility to use a single quantum
channel to perform arbitrary $n$-qubit QST, as shown schematically
in Fig.~\ref{F.NqubitQSTA}.

To achieve this goal, we now turn our attention to a detailed
analysis of the dynamics in the many-body regime.
Let's consider the spin model given in Eq.~\ref{E.HamBus}, with
uniform couplings~($J_i=J$, $\forall i$).
This model  can be
mapped to a spinless fermion model via a Jordan-Wigner
transformation~\cite{LiebSM1961}
\begin{equation}
\label{E.Hfermrealspace}
 \hat{\mathcal{H}}=-2\sum_{i{=}1}^N \hat{c}^{\dagger}_i \hat{c}_{i{+}1} + \mbox{h.c.}-\sum_{i=b_1,b_2} 2 h \hat{c}^{\dagger}_{i} \hat{c}_{i}~.
\end{equation}
where we have taken $J=1$ as our energy unit.
\antoC{Notice that} the
magnetic field is zero everywhere but at the barrier spins,
on sites $b_1=n+1$ and $b_2=N-n-1$, where it takes the same value
$h>0$.
The most general initial $n$-qubit pure state
reads~\cite{footnote}
\begin{eqnarray}
\label{E.nqstate0}
\nonumber
 \ket{\Psi(0)}_{12..n}&{=}& a_0\ket{0}{+}\!\!\!\!\sum_{n_1{=}1}^{n}a_{n_1}\ket{n_1}{+}\!\!\!\!\!\!\!\sum_{n_1{<}n_2{=}1}^{n}\!\!\!\!a_{n_1 n_2}\ket{n_1 n_2}{+}...
\\&&...{+}\sum_{n_1 < n_2 < \ldots < n_j{=}1}^{n}a_{n_1 n_2...n_j}\ket{n_1 n_2...n_j} +  \nonumber \\
&& \ldots + a_{n_1 n_2...n_n}\ket{n_1 n_2...n_n},
\end{eqnarray}
The last spin $n$ of the sender string is coupled to the first
barrier spin on site $n+1$, which is the first spin of the quantum
channel $\Gamma$, made out of $N-2n$ spins; the last spin of
$\Gamma$ (the second barrier) is coupled to the first spin of the
$n$ receiver string, located at the other end of the chain. In
Fig.~\ref{F.1qubitQSTMB} an instance of such a setting is shown
for $n=2$. Finally, both the spin bus $\Gamma$ and the receiver
spins are initialized in $\ket{0}$.
\antoC{We now} derive the time-evolved state of the whole system, sender
spins, quantum channel, and receiver spins. Because $
\hat{\mathcal{H}}$ commutes with the total number of excitations,
each $n$-fermions sector is an invariant subspace, and the analysis
can be performed separately in each subspace.
A lengthy but straightforward calculation shows that
\begin{eqnarray}\label{E.Nqstatet}
 &\ket{\Psi(t)}_{12..N}&{=}a_0\ket{0}{+}\sum_{n_1,m_1,k_1{=}1}^{N}a_{n_1}(t)D^{k_1}_{n_1}U_{k_1}^{m_1}\ket{m_1}\\
&{+}&\!\!\!\!\!\!\!\!\!\!\!\!\!\!\!\!\!\!\!\!\!\!\!\!\!\!\!
\sum_{n_1{<}n_2;k_1{<}k_2;n,m{=}1}^{N}a_{n_1 n_2}(t) D^{k_1 k_2}_{n_1 n_2} U_{k_1}^{m_1}U_{k_2}^{m_2}\ket{m_1 m_2}{+}...\nonumber\\
 &...{+}&\!\!\!\!\!\!\!\!\!\!\!\!\!\!\!\!\!\!\!\!\!\!\!\!
\sum_{n^\uparrow{=}1}^{N}\sum_{k^\uparrow{=}1}^{N}
\sum_{m_{i}{=}1}^{N} a_{n^\uparrow} D_{k_i^\uparrow}^{n^\uparrow}
U^{k_i^\uparrow}_{m_i}\ket{\{m_i^\uparrow\}}~.\nonumber
\end{eqnarray}
\antoC{Here} the arrow in $x^\uparrow$ indicates that the set of $x$'s are
ordered with the location label increasing from left to right,
i.e., $ x^\uparrow{\equiv}x_1{<}x_2{<}...{<}x_N$, whereas
$U^{k_i^\uparrow}_{m_i}{=}U^{k_1}_{m_1}...U^{k_N}_{m_N}$ and
$a_{n_1 n_2}(t){=}\prod_{i}e^{-itE_{k_i}}$.
$D_{k_1...k_r}^{n_1...n_r}$ is the
determinant of the minor made up by taking the $\{n_1,...,n_r\}$
rows and the $\{k_1,...,k_r\}$ columns of the matrix $U$\antoC{, which}
diagonalizes the $N \times N$ hamiltonian matrix given by
Eq.~\ref{E.Hfermrealspace}, in the single particle sector, while
$E_{k_i}$ denotes the $k_i$-th eigenvalue.
Finally, by tracing out all of the spins but the receivers, the
reduced density matrix $\rho^{(N-n+1,...,N)}(t)$, describing the spins located on sites
$m=N-n+1,...,N$ \antoC{and embodying} the QST target string,
 is obtained. Its fidelity
with the state given by Eq.~\ref{E.nqstate0} can be evaluated via
Eq.~\ref{E.Fidelity}.

\section{Results}\label{S.Results}

\begin{figure}[h]
\includegraphics[width=\columnwidth]{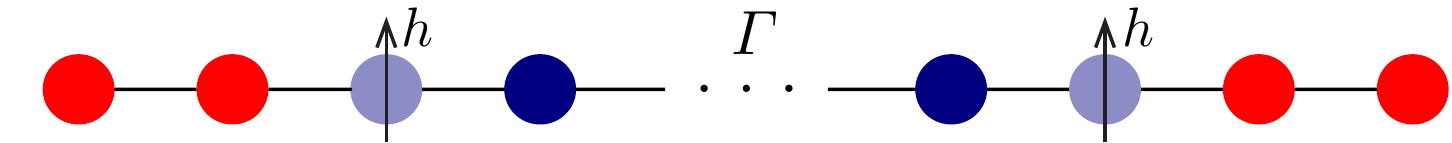}
 \caption{(Color online) 2-QST via Rabi-like mechanism between the two ends of the spin chain.}
  \label{F.1qubitQSTMB}
\end{figure}

In this Section we apply the above formalism to the first
non-trivial case, i.e., the transfer of a 2-qubit state residing
at sites 1 and 2, $\ket{\Psi(0)}_{12}=\alpha \ket{00}+\beta
\ket{01}+\gamma \ket{10}+\delta \ket{11}$, to the spins located at
the other end of the chain, $m=N-1,N$. To this \antoC{purpose,} we modify the
scheme depicted in Fig.~\ref{F.1qubitQSTA} by shifting the strong
magnetic field on qubits $3$ and $N-2$\antoC{, with the idea} to perform a
$2$-QST, as shown in Fig.~\ref{F.1qubitQSTMB}.

\noindent The reduced density matrix of the last two spins reads

\begin{strip}
\begin{equation}
\left(
\begin{array}{cccc}
 \mathcal{G}^{N-1}_{N} {\mathcal{G}^*}^{N-1}_{N} & {\mathcal{F}^*}_{N{-}1} \mathcal{G}^{N-1}_{N} & {\mathcal{F}^*}_{N} \mathcal{G}^{N-1}_{N} & {\alpha}^* \mathcal{G}^{N-1}_{N} \\
 \mathcal{F}_{N{-}1} {\mathcal{G}^*}^{N-1}_{N} & |\mathcal{G}^{m}_{N{-}1}|^2+|\mathcal{F}_{N{-}1}|^2 & \mathcal{G}^{m}_{N{-}1} {\mathcal{G}^*}^{m}_{N}+\mathcal{F}_{N{-}1} {\mathcal{F}^*}_{N} & {\mathcal{F}^*}_{m} \mathcal{G}^{m}_{N{-}1}+{\alpha}^* \mathcal{F}_{N{-}1} \\
 \mathcal{F}_{N} {\mathcal{G}^*}^{N-1}_{N} & \mathcal{G}^{m}_{N} {\mathcal{G}^*}^{m}_{N{-}1}+\mathcal{F}_{N} {\mathcal{F}^*}_{N{-}1} &|\mathcal{G}^{m}_{N}|^2+|\mathcal{F}_{N}|^2 & {\mathcal{F}^*}_{N} \mathcal{G}^{m}_{N}+{\alpha}^* \mathcal{F}_{N} \\
\alpha {\mathcal{G}^*}^{N-1}_{N} & \mathcal{F}_{m} {\mathcal{G}^*}^{m}_{N{-}1}+\alpha {\mathcal{F}^*}_{N{-}1} & \mathcal{F}_{m} {\mathcal{G}^*}^{m}_{N}+\alpha {\mathcal{F}^*}_{N} & 1-|\mathcal{G}^{m}_{N{-}1}|^2 -|\mathcal{G}^{m}_{N}|^2 -|\mathcal{F}_{N{-}1}|^2-|\mathcal{F}_{N}|^2\\
\end{array} \right)
\end{equation}
\end{strip}
\noindent Here the matrix elements
${\mathcal{F}_r{=}\beta \bra{r}U\ket{1}{+}\gamma
\bra{r}U\ket{2}}$ and $\mathcal{G}_s^r{=}\delta \bra{s,r}U\ket{1,2}$ depend
on the time-evolution operator $U{=}e^{-iHt}$
and the $m$-sum is intended to range between $1$ and $N{-}2$.

In order to obtain the average fidelity over all possible pure
initial states, \antoC{we conveniently adopt}  the following
parametrization for 2-qubit pure states
\begin{equation}
\ket{\Psi(0)}_{12}=\mathcal{R}_1\mathcal{R}_2\left(\sqrt{\frac{s{+}1}{2}}\ket{00}+\sqrt{\frac{s{-}1}{2}}\ket{11}\right)
\end{equation}
where $\mathcal{R}_{1,2}$ are local rotations acting on the first and
second spins, and $s$ characterize the initial amount of
entanglement~(as the concurrence is given by $C=\sqrt{1-s^2}$).
%
%
%
%

\subsection{Average Fidelity for general states}\label{sS.General}
\begin{figure}[h]
\includegraphics[width=0.98\columnwidth]{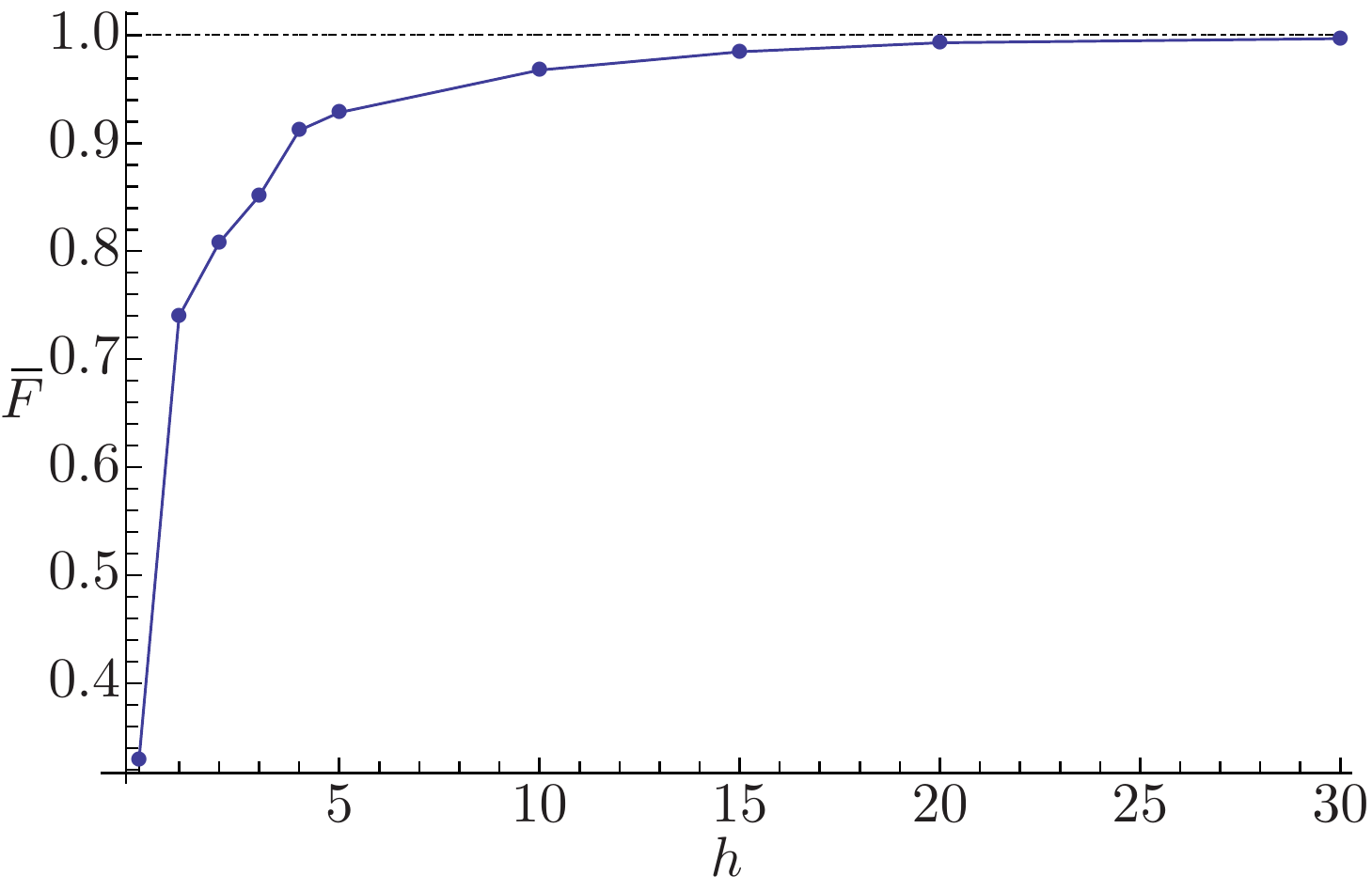}\\
\includegraphics[width=0.98\columnwidth]{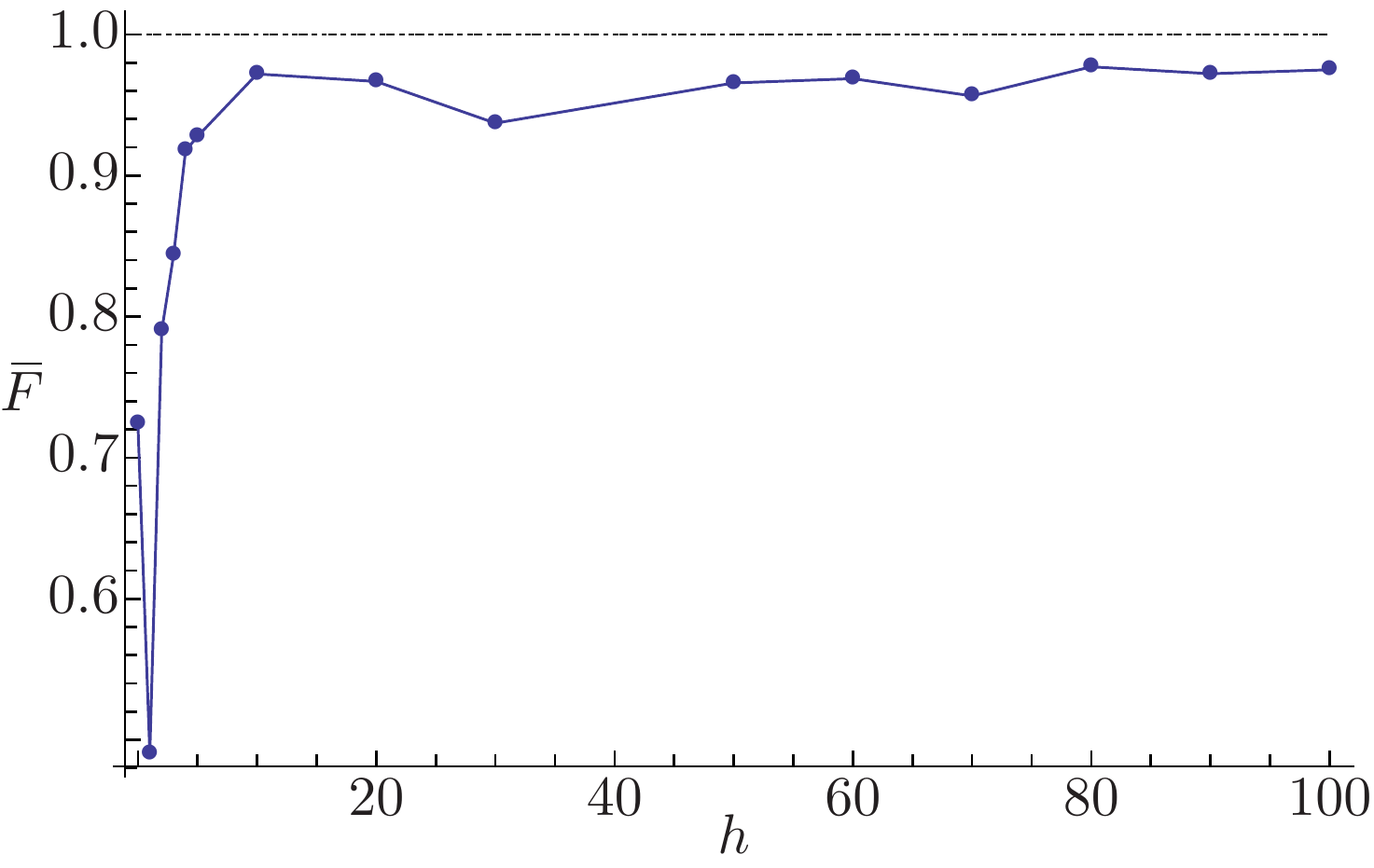}
\vskip -12pt
\caption{(Color online) Maximum average fidelity $\bar{F}$  for
the 2-QST \antoC{through chains made of $N{=}7$ and $N{=}8$ spins,
for time scans $t{\in} [0,2\times 10^4]$ and $t{\in} [0,6 \times 10^4]$,
respectively~(upper and lower
panels). Notice how} the presence of
strong magnetic fields greatly enhances the transfer efficiency
both for even and odd chains. Nevertheless, the protocol requires
weaker magnetic fields and less time in order to be performed on
odd-$N$ chains.\label{F.avF7F8}}
\end{figure}

With the reduced density matrix written above, we obtained an
analytic expression for the average fidelity $\bar{F}(t)$~(not
reported here for the sake of brevity). The results are shown in
Fig.~\ref{F.avF7F8}, where we report the maximum average fidelity
$\bar{F}(t)$ achieved at an optimal time $t^*$, scanned over the
time interval $[0,t_{max}]$, for different values of the magnetic
field $h$ on the barrier spins. The length of the quantum channels
chosen in the figures is a minimal one, that is $N=7$ and $N=8$,
in order to demonstrate a proof-of-principle of the setup.
Longer
chains will be addressed in the following subsection.
We see that the fidelity increases if larger and larger values
values of $h$ are taken, and that spin-buses with an odd number of
sites perform better that even-numbered ones, both because they
require smaller barrier magnetic fields and because the optimal
times where the maximum fidelity is achieved are shorter.

\subsection{Average Fidelity for subsets of states}
In the previous subsection we dealt with the transfer of the most
general two-qubit state, and the dynamics of the chain involved all of these
(invariant) subspaces\antoC{, in presence of $n=0,1,2$
excitations}.
\antoC{We now switch to cases where} the quantum state to
be transferred is not completely unknown, but it belongs to a known
subset. \antoC{We will see that t}he above results can be further improved and, in
addition, the analytical formula for the average fidelity
simplifies.
Indeed, for states of the form
$\ket{\Omega_1(0)}_{12}=b\ket{01}+c\ket{10}$ and
$\ket{\Omega_2(0)}_{12}=a\ket{00}+d\ket{11}$, the QST exploits the
invariance of the respective subspaces, and interference effects in
the average fidelity are strongly suppressed.

\begin{figure}[h]
\includegraphics[width=0.98\columnwidth]{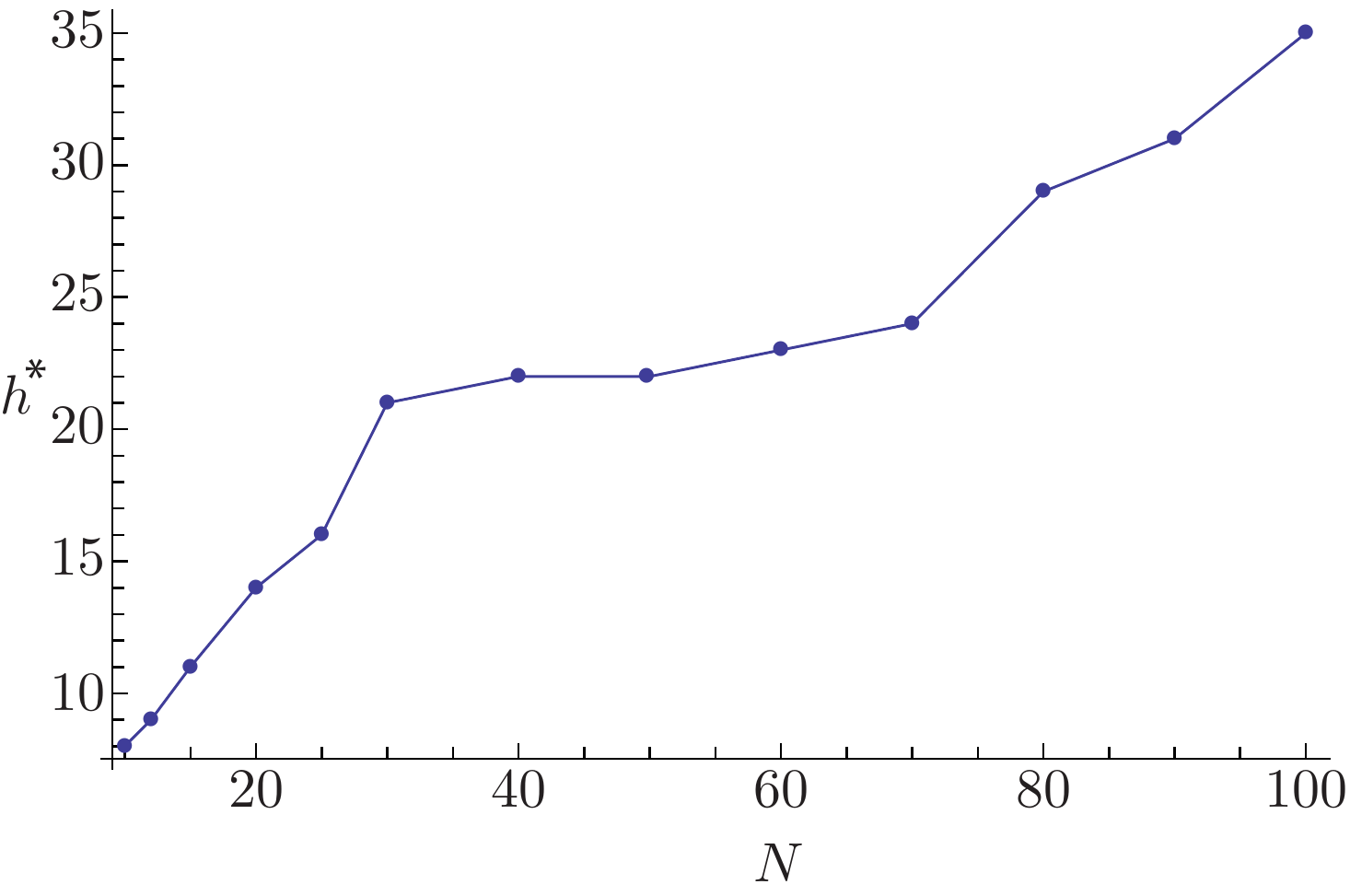}
\vskip -12pt
\caption{(Color online) Values of $h$ as a function of $N$
for which the average fidelity $\bar{F}$ of the 2-QST of states,
belonging to the subset $\ket{\Omega_1}$, is larger than $0.95$. The
time scan has been performed over the interval $t{\in} [0,1.3
\times 10^4]$.  \label{F.maxfid0110}}
\end{figure}

Restricting our considerations to states of the form $\ket{\Omega_1(0)}$, the average fidelity reads
\begin{eqnarray}\label{E.Fid0110}
\bar{F}(t)&{=}&\frac{1}{3}\left(| f_{N{-}1,1}|^2{+}| f_{N,2}|^2+\frac{| f_{N{-}1,2}|^2}{2}+\frac{| f_{N,1}|^2}{2}\right)\nonumber\\
&+&\frac{1}{3}Re\left[f_{N,2}f^*_{N{-}1,1}\right]
\end{eqnarray}
where $f_{j,i}$ denotes the transition's amplitude of the
excitation from site $i$ to site $j$.
In Figs.~\ref{F.maxfid0110}
we report the threshold value of the magnetic field $h^*$ as a
function of the chain's length $N$. \antoC{This} yields an average
fidelity, over the class of input states described by
$\ket{\Omega_1}$, larger \antoC{than} 0.95. It turns out that, also for
quite long chains, a high-quality 2-QST can be achieved with our
method provided that suitable values of $h$ are chosen.

\noindent Similar results are obtained by restricting the QST  to the states
$\ket{\Omega_2}$ (involving the 2-particle subspace). The average
fidelity\antoC{, in this case,} is
\begin{eqnarray}\label{E.Fid0011}
\bar{F}(t)&{=}&~\frac{1}{2}-\frac{1}{6}\sum_{n{=}1}^{N{-}2}\left(| g_{1,2}^{n,N{-}1}|^2+| g_{1,2}^{n,N}|^2\right)\nonumber\\
&+&\frac{1}{3}\left(| g_{1,2}^{N{-}1,N}|^2+Re\left[g_{1,2}^{N{-}1,N}\right]\right)~,
\end{eqnarray}
where $g_{1,2}^{r,s}$ is the transition's amplitude in the two excitations subspace.

Is is worthwhile to mention that the present work, as far as
we are aware of, relates for the first time
2-QST and transition amplitudes in a functional form via Eqs.~\ref{E.Fid0110} and
~\ref{E.Fid0011}\antoC{. This is achieved in the very same spirit as the framework leading to
Eq.~\ref{E.Fid01}, which relates functionally 1-QST and single
particle excitation transfer amplitudes. The results presented here, which are
valid for general quantum channels ruled by Eq.~\ref{E.HamBus},
may pave the way to investigate other than Rabi-like protocols for
the achievement of high-quality 2-QST.}
\\
\antoC{Finally,} we notice that the two subclasses $\ket{\Omega_1}$ and
$\ket{\Omega_2}$ span respectively the two-qubit pure states with
$\{0,1\}$- and $\{0,2\}$-spin flipped, respectively. This means
that the proposed scheme is able to transfer, to the other end of
the spin chain, outputs obtained by a magnetization-conserving
 unitary gate on two-qubits\antoC{. Examples of such a gate
 are the
Controlled-Z gate, (the target state of) a Fredkin gate, and
arbitrary single- and two-qubit phase gates.}

\section{Conclusions}\label{S.Conclusions}
In this paper we addressed the problem of the transfer of
many-body quantum states by means of open, finite 1D
spin-$\frac{1}{2}$ chains, interacting via $XX$-Heisenberg
exchange. Based on the protocol introduced in
Refs.~\cite{Lorenzoetal13} and \cite{Paganellietal13}, we proposed
 a setting capable to transfer arbitrary 2-qubit quantum
states between the end-points of a spin chain. The key element of
the protocol is the presence of a strong magnetic field applied on
two barrier qubits, on each side of the sender and receiver
strings connected to the spin bus. This magnetic field effectively
decouples the sender and receiver spins from the quantum channel,
and results in an effective hamiltonian supporting Rabi-like
oscillations of the excitations between the two ends of the spin
chain. We expressed the average fidelity in terms of the
corresponding excitation transfer amplitude and found that spin
chains made of an odd number of spins generally perform better
than even-numbered chains. By this we mean that the maximum average
fidelity obtained is higher at fixed intensities of the magnetic
field and the time duration of the protocol is shorter.

We also solved the dynamics for general $n$-qubit sender states,
with $n>2$, and this result may open the way to investigate
$n$-QST as well as distribution of multipartite entangled states,
relating in a functional form, amenable to theoretical
investigations, the amplitude for multiple excitation transfer and
$n$-QST fidelity. In this direction, we reported the explicit
formulas for 2-QST which may trigger the search and optimization
of other QST protocols.

\section{Acknowledgement}
{TJGA is supported by the European Commission, the
European Social Fund and the Region Calabria through the program POR
Calabria FSE 2007-2013 - Asse IV Capitale Umano-Obiettivo Operativo
M2. S.P.  acknowledges
partial support from MCTI and UFRN/MEC (Brazil).
G. L. G. acknowledges support from Compagnia di San Paolo.
}



\begin{thebibliography}{99}
\bibitem{footnote}
A multi-spin computational basis notation is here adopted, that is, the ket $\ket{n_1 n_2...n_i}$ denotes the positions along the chain of the flipped spins, e.g., $\ket{n_1{=}2\; n_2{=}4}=\ket{01010...}$. Note that, because of the increasing ordering, no phase factor due to the Jordan-Wigner transformations arises. Thus, provided the same increasing ordering is in use when referring to fermion excitations, the same notation indicates the presence of a Jordan-Wigner fermion at postion $n_i$.
\end{thebibliography}
\end{document}